\newcommand{\tabincell}[2]{\begin{tabular}{@{}#1@{}}#2\end{tabular}}

\newcommand{\phead}[1]{\vspace{1mm} \noindent {\bf #1}}
\newcommand{\wei}[1]{\textcolor{blue}{{\it [Wei says: #1]}}}
\newcommand{\peter}[1]{\textcolor{red}{{\it [Peter says: #1]}}}

\newcommand{\tool}{{{GUIWatcher}}\xspace}

\newcommand{\rqbox}[1]{\begin{tcolorbox}[left=4pt,right=4pt,top=4pt,bottom=4pt,colback=gray!5,colframe=gray!40!black,before skip=2pt,after skip=2pt]#1\end{tcolorbox}}

\documentclass[10pt,conference]{IEEEtran}
\IEEEoverridecommandlockouts

\usepackage{xcolor}
\usepackage{amsmath,amssymb,amsfonts}
\usepackage{algorithmic}
\usepackage[pdftex]{graphicx}
\usepackage{grffile}
\usepackage{textcomp}
\usepackage{xcolor}
\usepackage[numbers]{natbib}
\usepackage{subcaption}
\usepackage{xspace}
\usepackage{booktabs, tabularx}
\usepackage{url}
\usepackage{multirow}
\usepackage[colorlinks=false, linkcolor=blue, citecolor=blue, bookmarks=true,pagebackref=false]{hyperref}
\usepackage{tcolorbox}
\usepackage{verbatim}
\usepackage{balance}
\usepackage{float}
\usepackage[T1]{fontenc}

\definecolor{lightblue}{RGB}{135, 206, 250}
\begin{document}

\title{\tool: Automatically Detecting GUI Lags by Analyzing Mobile Application Screencasts
}

\author{
    \IEEEauthorblockN{Wei Liu\textsuperscript{1}, Feng Lin\textsuperscript{1}, Linqiang~Guo\textsuperscript{1}, Tse-Hsun (Peter) Chen\textsuperscript{1}, Ahmed E. Hassan\textsuperscript{2}}
    \IEEEauthorblockA{\textit{\textsuperscript{1}Software PErformance, Analysis, and Reliability (SPEAR) lab, Concordia University, Montreal, Canada}}
    \IEEEauthorblockA{\textit{\textsuperscript{2}Software Analysis and Intelligence Lab (SAIL), Queen’s University}}

    \IEEEauthorblockA{w\_liu201@encs.concordia.ca, feng.lin@mail.concordia.ca, g\_linqia@live.concordia.ca}
    \IEEEauthorblockA{peterc@encs.concordia.ca, ahmed@cs.queensu.ca}
}
\maketitle

\begin{abstract}
The Graphical User Interface (GUI) plays a central role in mobile applications, directly affecting usability and user satisfaction. Poor GUI performance, such as lag or unresponsiveness, can lead to negative user experience and decreased mobile application (app) ratings. In this paper, we present \tool, a framework designed to detect GUI lags by analyzing screencasts recorded during mobile app testing. \tool uses computer vision techniques to identify three types of lag-inducing frames (i.e., janky frames, long loading frames, and frozen frames) and prioritizes the most severe ones that significantly impact user experience. Our approach was evaluated using real-world mobile application tests, achieving high accuracy in detecting GUI lags in screencasts, with an average precision of 0.91 and recall of 0.96. The comprehensive bug reports generated from the lags detected by \tool help developers focus on the more critical issues and debug them efficiently.
Additionally, \tool has been deployed in a real-world production environment, continuously monitoring app performance and successfully identifying critical GUI performance issues. By offering a practical solution for identifying and addressing GUI lags, \tool contributes to enhancing user satisfaction and the overall quality of mobile apps.
\end{abstract}


\section{Introduction}
\label{sec:introduction}

Non-functional performance characteristics, particularly the responsiveness and smoothness of the Graphical User Interface (GUI), are critical to user satisfaction with mobile applications~\cite{2015_IEEE_Software_Mobile_App_Users_Complain, 2018_WWW_Rating_for_App_Reviews, 2022_JSS_What_factors_affect_the_UX_in_mobile_apps}. As mobile devices have become more powerful and widely used, user expectations for seamless performance have risen accordingly. 
Performance issues like GUI lags can significantly and directly impact users' perceptions of an app's quality. Frequent or severe lags may cause users to abandon the app and leave extremely negative reviews, ultimately damaging the app's ratings and overall reputation. 

Despite this growing emphasis on performance, conventional GUI testing tools often fall short in detecting issues that align with users' perceptions of lag. While some studies employ static analysis of source code to identify performance issues in mobile applications~\cite{2014_ICSE_Characterizing_and_detecting_performance_bugs, 2019_EMSE_iPerfDetector,2019_SANER_Characterizing_and_Detecting_Inefficient_Image_Displaying_Issues, 2020_EMSE_statically_detectable_performance_issues, 2021_ICSE_IMGDroid_Detecting_Image_Loading_Defects, 2023_ASE_Detection_Thread_Misuses}, the issues identified by these methods may not accurately reflect users' experiences of unresponsiveness. Additionally, static analysis tools depend heavily on known performance patterns, limiting their applicability in real-world scenarios. Other traditional approaches commonly focus on low-level metrics like CPU and memory usage, which do not necessarily correlate with perceived performance issues~\cite{2022_IST_resource_influences_UI_responsiveness, adb, perfetto}. 


Screencasts are recordings of a mobile device’s screen during app usage or GUI testing. Analyzing screencasts offers a direct way to identify potential GUI performance issues, such as lag, from the user's perspective. These screencasts capture the exact sequence of individual frames rendered by the mobile system, with each frame representing a static UI image at a specific time point marked by a timestamp. This creates a rich data source for performance analysis. 
However, manually reviewing screencasts to determine whether the app is slow or experiencing lagging is both time-consuming and labor-intensive. For a one-minute screencast recorded at 60 Hz, reviewers must examine 3,600 frames individually to pinpoint the issues for further analysis. This process becomes even more infeasible in a Continuous Integration (CI) setting, where frequent code updates and automated testing produce a large volume of screencasts.  

To address these challenges, we collaborated with Company A to develop \tool, a framework that uses computer vision techniques to automatically detect GUI lags from mobile application screencasts---lags that users perceive as performance delays or a lack of smoothness during app usage.
These screencasts provide direct insight into how applications behave from the end user’s perspective, making them ideal for identifying performance issues. While minor lags might be acceptable from a company’s standpoint, severe lags can significantly harm user experience. \tool further identifies these critical GUI lags, enabling companies to address high-impact issues before they affect a broader user base.

\tool is a framework composed of two key components: (1) \textit{GUI Testing \& Video Capture} and (2) \textit{Detection of GUI Performance lags}. In the first component, we conduct GUI tests on mobile applications to simulate user interactions, such as tapping. During these tests, we record screencasts that capture the mobile device's screen. To minimize additional performance overhead during testing and recording, we use external hardware to record the screencast efficiently. 
In the second component, we implement a rule-based approach that combines image similarity and object detection techniques to identify three types of GUI lags: (1) \textit{Janky Frames}---dropped frames that disrupt the smoothness of the UI, (2) \textit{Long Loading Frames}---frames that indicate long resource loading, causing bad user experience, and (3) \textit{Frozen Frames}---frames where the UI appears unresponsive, giving the impression of a freeze in user interaction. 

We begin by applying the image similarity techniques of Structural Similarity Index Measure (SSIM)~\cite{2004_SSIM_Transactions_on_Image_Processing} to the screencast to extract static frames (i.e., frames that do not change) to narrow the analysis scope and improve efficiency. Since users perceive these static frames as unchanged for an extended period, they may create a sense of lag and indicate potential performance issues.
Next, we classify the static frames into potentially problematic frames that may signify GUI lags. To assess these frames, we calculate their duration by measuring the intervals between their presentation times. If any interval exceeds 100 milliseconds (set based on HCI studies~\cite{1968_AFIPS_Response_time_in_man_computer, 1994_Usability_Engineering, zippy_Android_apps, 2015_MOBILESoft_performance_parameters}), we classify it as a GUI lag caused by janky frames. We then distinguish long loading frames from frozen frames based on their content. Long loading frames typically contain a placeholder representing content that has yet to be loaded, serving a visual cue to users that loading is in progress. 

To differentiate between these frame types, we fine-tune a pre-trained YOLO~\cite{yolov} object detection model using a dataset of images we collected and manually labeled for any placeholders. Using the fine-tuned YOLO model, we identify placeholders within the frames. If placeholders are detected and the total duration exceeds one second, we classify it as a GUI lag caused by long loading frames. Conversely, if no placeholders are detected and the total duration exceeds 100 milliseconds, we classify it as a GUI lag caused by frozen frames. 
By combining these techniques with predefined rules based on thresholds from prior HCI studies~\cite{1968_AFIPS_Response_time_in_man_computer, 1994_Usability_Engineering, zippy_Android_apps, 2015_MOBILESoft_performance_parameters}, \tool effectively detects the three types of GUI lags. Furthermore, \tool distinguishes severe GUI lags based on the duration of these problematic frames, highlighting those that significantly degrade the user experience.

To evaluate the effectiveness of \tool, we conducted experiments using a dataset of real-world testing results, specifically mobile screencasts collected from the Continuous Integration (CI) test runs at company A. 
The evaluation results show that \tool achieves an average precision of 0.91 and recall of 0.96 in detecting GUI lags. 
Furthermore, we deployed \tool in a real-world production environment, which is part of the CI to analyze screencasts to monitor app performance. \tool has successfully identified many severe GUI lags. Positive feedback confirms that the comprehensive bug reports generated from these lags provide developers with actionable insights for debugging and locating performance issues. The reports also help them prioritize fixing critical issues that have a larger impact on user experiences. 

The main contributions of this paper are as follows: 
\begin{itemize}
	\item We present \tool, a GUI performance testing framework that effectively detects user-centric GUI lags by analyzing mobile screencasts. 
 \item  \tool analyzes various factors, such as user interactions (e.g., lags during scrolling can be more impactful to human perception), duration and frequency of lags, and visual context (e.g., GUI elements experience lags), to help developers prioritize performance issues that directly impact user experience.

	\item We evaluate \tool using real-world testing results, achieving an average precision of 0.91 and recall of 0.96 in detecting GUI lags.


	\item We deployed and integrated \tool into the CI pipeline at company A to continuously monitor the GUI performance of mobile applications. We also discussed the positive feedback we received. 

\end{itemize}

In short, we hope our successful industry collaboration can inspire practitioners and researchers working on this direction. Future studies may consider adopting more user-centered approaches in their performance testing processes. 

\phead{Paper organization.} 
Section~\ref{sec:background} provides background on mobile screencasts and outlines three types of GUI lags. Section~\ref{sec:approach} details our approach.
Section~\ref{sec:evaluation} evaluates our approach using real-world testing results, while Section~\ref{sec:discussion} discusses its production impact. Section~\ref{sec:related} discusses related work and Section~\ref{sec:conclusion} concludes the paper.

\section{Background}
\label{sec:background}
In this section, we first provide background information on mobile screencasts that capture the rendered GUI as videos. 
Next, we discuss the details of three types of GUI lags and how we determine if a lag affects the user-perceived quality.  

\begin{figure*}
	\centering
    \includegraphics[width=0.9\textwidth]{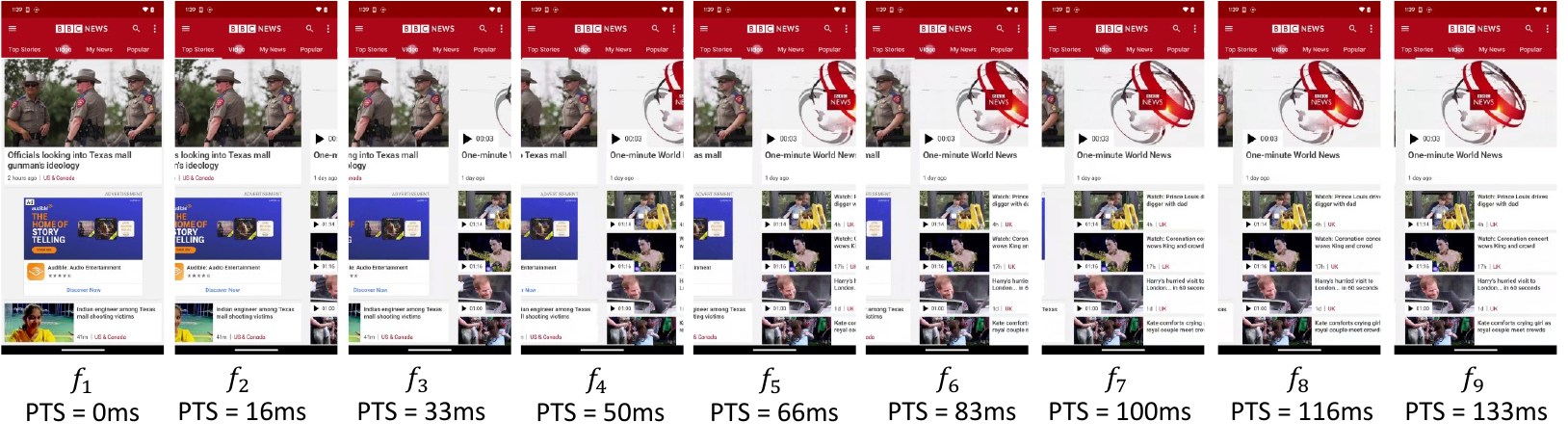}
	\caption{A sequence of frames in a recorded screencast, with each frame annotated by index (e.g, $f_{1}$) and PTS.}
	\label{fig_normal_frame}
\end{figure*}

\subsection{Mobile Testing and Screencasts} 
Users expect mobile applications to be smooth and responsive, as performance issues, such as slow loading times or an unresponsive GUI, can significantly degrade the user experience, leading to low app ratings or even uninstallation~\cite{2022_JSS_What_factors_affect_the_UX_in_mobile_apps,2022_TSE_Survey_of_Performance_Optimization}. Consequently, GUI-based mobile testing is essential for ensuring app quality as it reflects actual app usage. Developers typically write test scripts or utilize automated testing tools, such as Monkey~\cite{Monkey}, to simulate user interactions and mimic app usage. After test execution, developers analyze the performance metrics (e.g., CPU usage and memory consumption) and running logs collected during testing to identify potential performance issues. However, these metrics often fail to capture the user’s real experience. For instance, high CPU utilization does not always correlate with perceivable lags, and logs may overlook visual issues. In contrast, screencasts provide a visual recording of the app’s behavior, allowing for a more comprehensive user experience analysis. By analyzing screencasts recorded during automated tests, we can better uncover user-perceived performance issues. 

A mobile screencast is a recording that captures the behavior of applications on a mobile device’s screen in response to a user's actions (e.g., tapping). It consists of a sequence of individual frames, each representing a static UI image at a specific time point, marked by a timestamp. 
Screencasts are often recorded at the same frame rate as the device~\cite{screen_record_frame_rate}. Each time the screen updates, a new frame is captured. For instance, mobile devices typically operate at a standard frame rate of 60 Frames Per Second (FPS)~\cite{Android_frame_rate}, which means that the screencast is also recorded at 60 FPS, with approximately 16.7 milliseconds~\cite{Android_Performance_Patterns} between consecutive frames. 
Therefore, screencast playbacks reflect what users see on the mobile device, and analyzing the screencast can help identify performance issues (e.g., lags) that occur during app usage.

Figure~\ref{fig_normal_frame} illustrates a screencast composed of 9 frames, each annotated with a timestamp indicating its Presentation Time Stamp (PTS). For example, frame 1 ($f_1$) is shown at timestamp 0 ms, followed by frame 2 ($f_2$) at timestamp 16 ms, and so on, creating a seamless transition that users perceive as smooth application performance. However, if these transitions are not rendered smoothly, they can result in noticeable GUI lags. 

\subsection{Three Types of GUI Lags that Indicate Possible Performance Issues} 
In our collaboration with Company A, we uncovered three distinct types of GUI lags that significantly impacted the user experience: janky frames, long loading frames, and frozen frames. 
These performance issues reflect GUI lags and negatively impact user-perceived app quality. 
Every GUI lag is annotated as an interval $[f_{start}, f_{end}]$, where $f_{start}$ and $f_{end}$ denote the start and end frame indices of the lag within the screencast (i.e., a sequence of frames). 
For the lag to be perceptible to end-users, the frames before and after the interval need to exhibit continuous transition, which allows users to perceive the lag in the transition frames. Below, we discuss the three types of GUI lags that we studied. 

\begin{figure}
	\centering
	\begin{subfigure}[b]{0.45\textwidth}
		\centering
		\includegraphics[width=\textwidth]{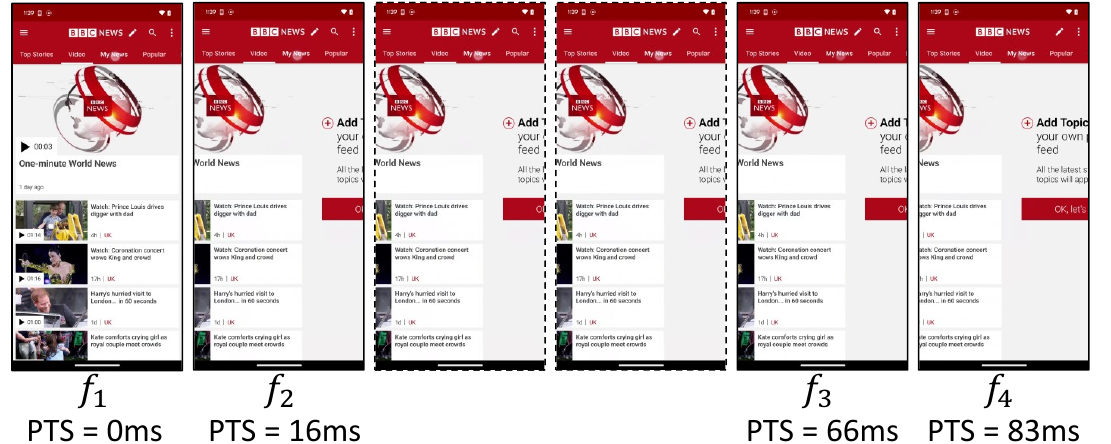}	
		\caption{\texttt{Janky frames between $[f_{2}, f_{3}]$}}
        \label{fig:GUI_perf_jank}
	\end{subfigure}

    \begin{subfigure}[b]{0.45\textwidth}
		\includegraphics[width=\textwidth]{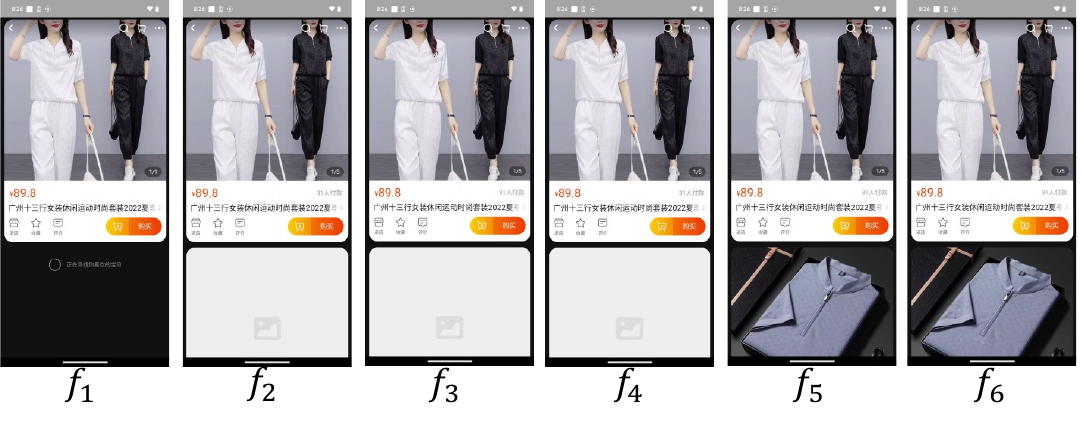}
		\caption{\texttt{Long loading frames between $[f_{2}, f_{4}]$}}
        \label{fig:GUI_perf_load}
	\end{subfigure}

     \begin{subfigure}[b]{0.45\textwidth}
		\includegraphics[width=\textwidth]{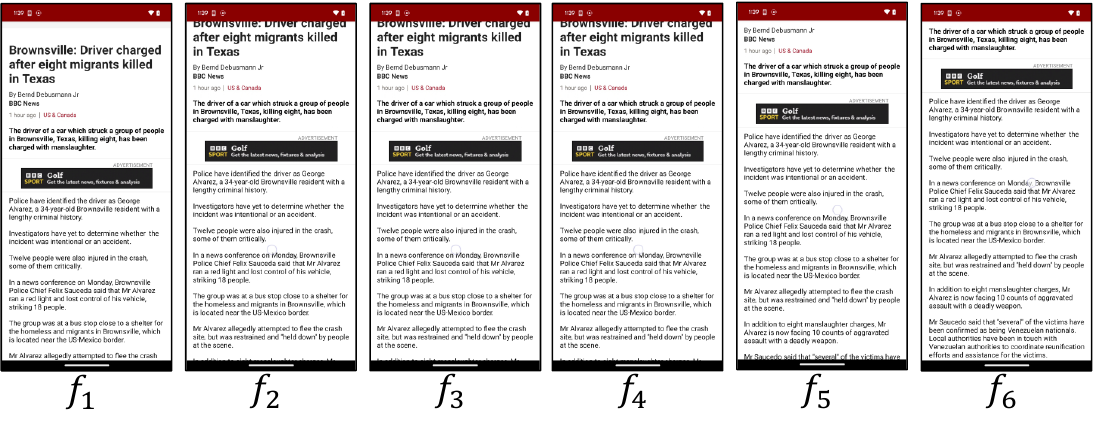}
		\caption{\texttt{Frozen frames between $[f_{2}, f_{4}]$}}
        \label{fig:GUI_perf_frozen}
	\end{subfigure}

    \caption{Three types of GUI lags.}
    \vspace{-3mm}
	\label{fig:GUI_perf_lags}
\end{figure}

\subsubsection{Janky Frames} 
Mobile OSes like Android render the UI by generating and displaying frames on the screen. Ideally, a mobile app should run at a consistent frame rate of 60 Frames Per Second (FPS). However, if the app experiences slow UI rendering, the system may be forced to skip some frames, causing a recurring flicker on the screen known as jank~\cite{Developers_UI_jank_detection}. This jank can lead to an unstable frame rate and increased latency~\cite{Perfetto_Android_Jank_detection_with_FrameTimeline}, resulting in a bad user experience. 
For instance, as illustrated in Figure~\ref{fig:GUI_perf_jank}, the Android system renders a sequence of frames. After rendering frame 2, the system skips or delays the rendering of subsequent frames. Consequently, at the 33ms and 50ms timestamps, the content of frame 2 remains displayed on the screen, as no new frames are rendered (indicated by the images surrounded by dotted lines). Frame 3 is only rendered and displayed on the screen until the timestamp reaches 66ms. As a result, users may perceive the screen as stuck between 16ms and 66ms timestamps, as the screen display does not change, leading to a noticeable lag. Thus, the frames $[f_{2}, f_{3}]$ are identified as janky frames.

\subsubsection{Long Loading Frames}
Mobile apps need to load essential resources, such as images, to display various content and enhance the user experience. However, if loading and displaying take too long, end-users may experience more challenges as page resources load gradually~\cite{speed_matter}. 
For example, inefficient or improper image loading and display operations can severely impact app performance~\cite{2019_SANER_Characterizing_and_Detecting_Inefficient_Image_Displaying_Issues, 2021_ICSE_IMGDroid_Detecting_Image_Loading_Defects}, potentially leading to long loading frames that affect the user-perceived quality of the app. As shown in Figure~\ref{fig:GUI_perf_load}, an image placeholder begins to appear in frame $f_{2}$ and is not fully filled with an image until frame $f_{5}$, exhibiting three consecutive loading frames. If the number of consecutive loading frames is large, the user may feel the app is unresponsive. 

\subsubsection{Frozen Frames}
Responsiveness is a fundamental metric that significantly impacts user experience~\cite{2022_TSE_Survey_of_Performance_Optimization}. However, certain performance anti-patterns, such as accessing data on the UI thread, can cause screen frames to appear sluggish or frozen, adversely affecting users' perception of the app’s performance~\cite{2019_EMSE_iPerfDetector}. These data access operations can be computation-intensive and time-consuming, making the UI unresponsive until the data is retrieved, causing frozen frames. For example, in Figure~\ref{fig:GUI_perf_frozen}, the page scrolls accordingly when the user scrolls up on the screen. However, the frames $[f_{2}, f_{4}]$ remain frozen during this transition, which negatively impacts the user experience and reflects potential performance issues. 



\subsection{Determining if GUI Lags Affect the User-perceived Quality} 
Assessing whether GUI lags impact the user-perceived quality of an app is crucial for maintaining a smooth, responsive experience that satisfies users. However, not all GUI lags are equally noticeable or problematic. Short and minor lags often go unnoticed and do not affect how users perceive the quality of an app~\cite{2015_High_Performance_Android}. Therefore, determining the threshold at which lags become perceptible and affect user satisfaction is key to prioritizing performance improvements.

Research in Human-Computer Interaction (HCI) and user experience design shows that UI feedback should occur within 100ms (approximately six frames at a 60Hz refresh rate) to ensure a smooth experience~\cite{1968_AFIPS_Response_time_in_man_computer, 1994_Usability_Engineering, zippy_Android_apps, 2015_MOBILESoft_performance_parameters}. Therefore, we follow the guideline and set 100ms as the threshold for identifying GUI lags caused by \textit{janky frames} or \textit{frozen frames}. On the other hand, the image loading time in mobile applications can vary significantly depending on factors such as image size, image format, network situation, and other related variables. Hence, we set the threshold for \textit{long loading frames} to one second. 

\section{Approach}
\label{sec:approach}

\begin{figure*}
	\centering
    \includegraphics[width=0.95\linewidth]{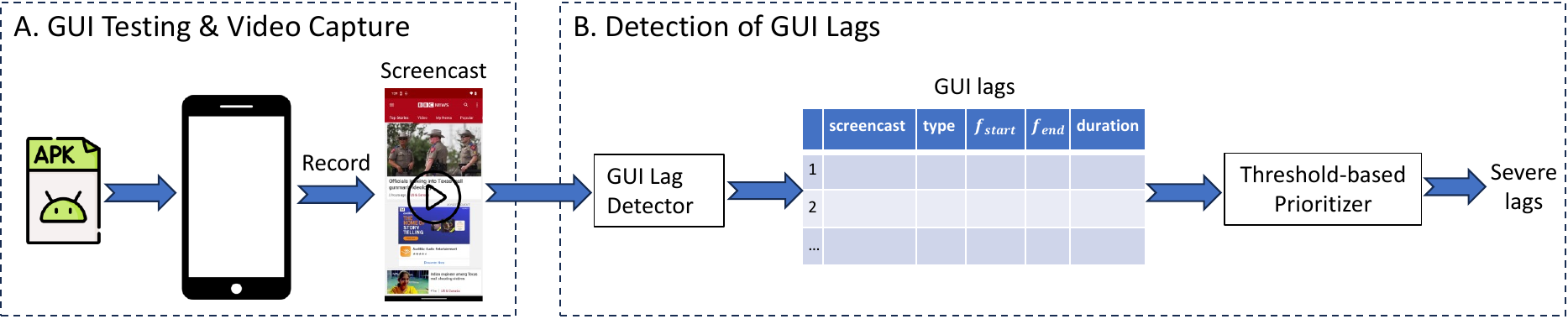}
	\caption{The overall architecture of \tool.}
	\label{fig_overview_approach}
\end{figure*}

Manually analyzing screencasts to identify GUI lags is a time-consuming and labor-intensive process that becomes nearly impossible as the complexity and number of mobile apps grow. With apps typically generating thousands of frames in a one-minute interaction, manually reviewing these frames to detect performance issues, such as janky, long loading, and frozen frames, is inefficient and error-prone. 

To address these challenges, we collaborated with our industry partner, Company A, to propose an automated approach for analyzing screencasts to detect GUI lags efficiently and accurately. Figure~\ref{fig_overview_approach} presents the overall architecture of the framework based on our approach, \tool.
It comprises two main components: \textit{GUI Testing \& Video Capture} and \textit{Detection of GUI lags}. 
\tool has been fully integrated into Company A's development workflow as part of the continuous testing and quality assurance processes. The CI pipeline triggers \tool every time the automated tests finish executing to generate a GUI lag detection report. To date, \tool has successfully detected and helped resolve many performance issues. 

Below, we discuss \tool in detail.

\subsection{GUI Testing \& Video Capture}
As illustrated in Figure~\ref{fig_overview_approach}, our approach utilizes GUI screencasts (i.e., a sequence of frames) as input for detecting GUI lags. The process begins with capturing these GUI screencasts from real mobile applications during GUI testing. During test execution, the tests automatically trigger the screen recording process. We record the screencasts at full Frames Per Second (FPS) to comprehensively analyze every screen change in response to user operations (e.g., tapping). To minimize the performance overhead during the recording process, we use specialized external hardware that captures and saves each frame as the mobile device renders it to the screen. Using such hardware minimizes the performance impact of screen recording and ensures that the recorded screencasts accurately reflect what human eyes would see. Once recorded, the screencasts are stored as video files for subsequent analysis. 

\subsection{Detection of GUI Lags}
\subsubsection{Lag Detector}
This phase aims to automatically detect GUI lags from screencasts. The output includes detailed information about the detected GUI lags, such as the screencast name, the type of GUI lags, the start and end frame indices of the lag within the screencast, and the duration of the lag-inducing frames. 


\begin{figure}
	\centering
    \includegraphics[width=0.98\linewidth]{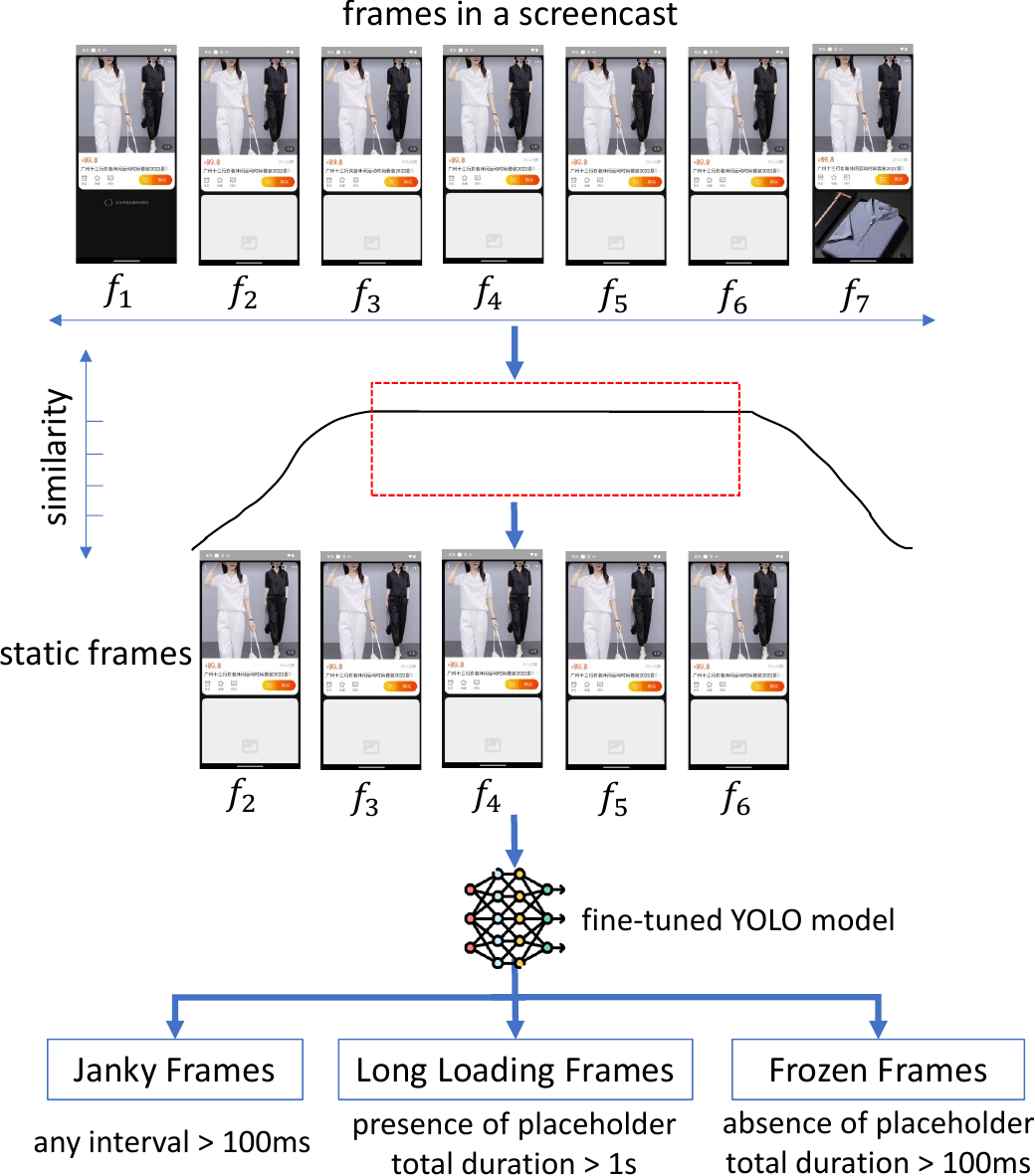}
	\caption{An illustration of GUI lag detection.}
	\label{fig_lag_detection}
 \vspace{-3mm}
\end{figure}

As illustrated in Figure~\ref{fig_lag_detection}, we first extract the static frames (i.e., frames that end users perceive as unchanged for some time) from the given screencast, which may or may not potentially be GUI lags. This approach helps focus the subsequent analysis on specific time frames of the screencast to speed up analysis. 
Inspired by prior studies~\cite{2023_UIST_Video2Action, 2023_ICSE_Efficiency_Matters_Speeding_Up_Automated_Testing_with_GUI_Rendering_Inference}, we employ the Structural Similarity Index Measure (SSIM)~\cite{2004_SSIM_Transactions_on_Image_Processing}, a highly effective image similarity technique that reflects human perception, to segment the screencasts and identify static frames. For every frame in the screencast, we calculate the similarity score relative to the previous frame. If the similarity score exceeds a predefined threshold, it indicates no human perceivable change compared to the prior frame. 
As an example, in Figure~\ref{fig_lag_detection}, frame $f_{3}$ to $f_{6}$ have a similarity score that exceeds the predefined threshold, indicating there are static frames from $f_{2}$ to $f_{6}$ (frame $f_{2}$ is included because it is the first frame that shows no subsequent change).

After extracting these static frames, we apply rule-based performance bug detection approaches to identify the three types of GUI lags discussed in Section~\ref{sec:background}. 

\phead{Janky Frames.} Jank occurs when frames are skipped or delayed, causing the screen to display the content of a previous frame. This makes users perceive the screen as momentarily stuck, as the display fails to update as expected. 
We detect jank by identifying frames where the interval between two rendered frames exceeds 16.7 milliseconds, corresponding to a standard frame rate of 60 frames per second (FPS). To implement this detection, we traverse all static frames to retrieve their presentation times and calculate the interval between consecutive frames by subtracting their presentation times. If any interval exceeds 16.7 milliseconds, a jank is detected. Additionally, if any interval exceeds 100 milliseconds, indicating the skipping of six frames (as discussed in Section~\ref{sec:background}), we classify it as a GUI lag caused by janky frames. The indices of the consecutive frames are marked as the start and end frames of the janky frames, annotated as an interval $[f_{start}, f_{end}]$.

\begin{figure}
	\centering
    \includegraphics[width=0.98\linewidth]{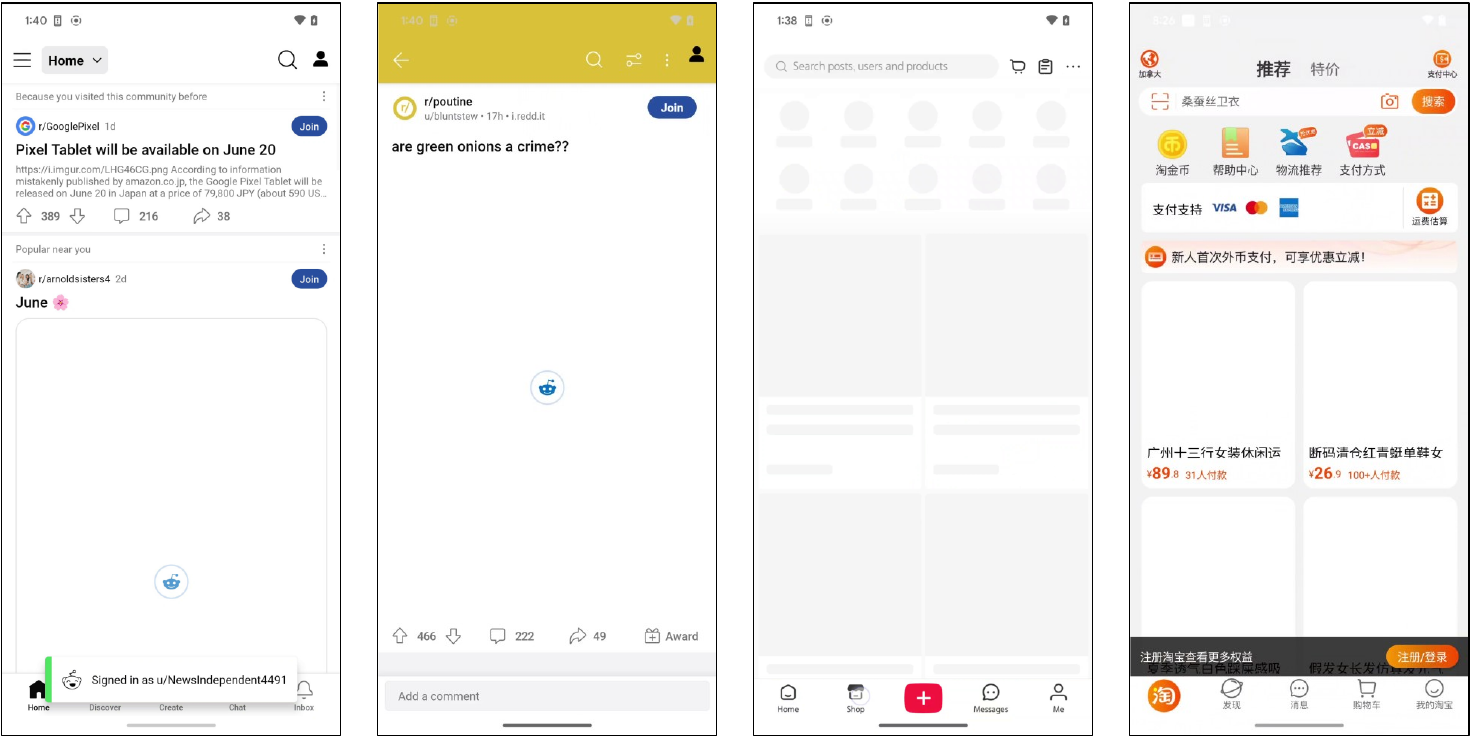}
	\caption{Examples of frames containing placeholders.}
	\label{fig_placeholder}
 \vspace{-3mm}
\end{figure}

\phead{Long Loading Frames.} Long loading frames indicate resource loading, which can significantly disrupt the user experience by causing noticeable delays. Typically, a placeholder in the frame represents content that has yet to be loaded, serving a visual cue to users that loading is in progress. Figure~\ref{fig_placeholder} presents examples of frames containing placeholders. To identify these loading frames, we use a fine-tuned object detection model to detect placeholders within the frames. We first build a dataset of images by collecting numerous mobile screenshots (i.e., images) and manually labeling any placeholders. Then, we fine-tune the pre-trained YOLO~\cite{yolov} model using this dataset. 
YOLO is a computer vision-based object detection model pre-trained with images from various sources, such as ImageNet~\cite{ImageNet}. However, its training data does not involve any context information about mobile app screens. The fine-tuning process optimizes the model's parameters with our mobile app data, enabling it to accurately detect the existence and location of placeholders in images. We evaluate the fine-tuned YOLO model and find that it achieves high precision and recall in detecting the placeholders. 

Specifically, we focus on the first frame of the static frames (i.e., consecutive frames with SSIM scores larger than a threshold). If a placeholder is detected by the fine-tuned YOLO model on the first frame and the duration between the first frame and the last frame exceeds one second (calculated by subtracting their respective presentation times), we classify this scenario as a GUI lag caused by long loading frames. The indices of the first and last frame in the static frames are marked as the start and end frames of the long loading frames, denoted as the interval $[f_{start}, f_{end}]$.

\phead{Frozen Frames.} Frozen frames occur when the content displayed on the screen becomes unresponsive, resulting in a stop where no changes happen on the screen. Unlike long loading frames, frozen frames do not contain explicit placeholders indicating resource loading. In contrast to janky frames, which happen when the frame rate drops inconsistently due to app overload, frozen frames represent a loss of responsiveness caused by a hang that temporarily locks up the app. We want to categorize the detected GUI lags into the corresponding type for better bug triage. Hence, to identify frozen frames, we employ the same fine-tuned YOLO~\cite{yolov} model to check for the presence of placeholders in the first frame of the static frames. If the placeholder is absent and the duration between the first frame and last frame exceeds 100 milliseconds (calculated by subtracting their respective presentation times), we classify these static frames as frozen frames. The indices of the first and last frame in the static frames are marked as the start and end frames of the frozen frames, denoted as the interval $[f_{start}, f_{end}]$.

\subsubsection{Prioritizing Severe GUI Lags}
Our approach can detect numerous instances of GUI lags. However, one issue is that some lags may be short (e.g., at the boundary of the 100ms threshold), while some may be long and significantly impact end users’ experiences. Moreover, the severe lags are more likely to indicate underlying performance issues.  
For instance, consider a scenario where a user clicks a button in a mobile application, but the screen transition is delayed by several seconds. Such a delay could signal inefficiencies in the rendering process or problems within the app’s underlying logic, both of which are key performance-related issues affecting the GUI. We categorize these prolonged lags as ``severe GUI lag'' as they significantly degrade the user experience and might indicate deeper performance bottlenecks that require immediate attention.


We employ a threshold-based approach to prioritize severe GUI lags because it provides a simple yet effective method for distinguishing between minor and major delays. 
The lag duration is the key feature influencing user perception of GUI performance. Therefore, we establish duration thresholds to classify any GUI lag exceeding these limits as severe. 
After consulting with user experience experts at Company A, we set three duration thresholds: $t_{janky}$,  $t_{load}$, and $t_{frozen}$, which correspond to janky frames, long loading frames, and frozen frames, respectively.
Any detected GUI lag exceeding these thresholds is considered severe lag by \tool and will undergo in-depth investigation. Although we cannot share the exact values of these duration thresholds due to the Non-Disclosure Agreement (NDA), practitioners willing to implement and deploy \tool can set their thresholds according to their specific requirements. 

\subsection{Efficiency of \tool and Recording Overhead}
\label{sec:implementation}

To minimize the overhead associated with screen recording on the mobile devices under test, we utilize external hardware for recording in controlled settings.
Due to the NDA, we cannot disclose the implementation details of the recording process. 
Nevertheless, to measure the recording overhead, we monitor performance metrics (e.g., response time, CPU utilization, memory consumption, and frame rates) to compare the performance before and after integrating the hardware for screen recording. After the integration, we observed no degradation in performance, as the performance metrics remained nearly the same (at most 10ms or so difference in response time). This indicates that the overhead is technically not human-observable, and analyzing the screencasts can accurately reflect users' perceptions of performance and GUI lags.

\section{Evaluation}
\label{sec:evaluation}

\subsection{Experimental setup}
\label{sec:setup}
\subsubsection{Collecting Dataset from Real-World Testing Results} 

To effectively detect and analyze GUI performance issues, we have deployed our tool, \tool, in a real-world production environment within the Continuous Integration (CI) pipeline of Company A. This integration enables automated data collection from CI test runs, which are executed regularly as part of the company’s quality assurance process. During each test run, \tool captures screencasts of the mobile application under various usage scenarios, simulating real user interactions such as tapping, scrolling, and navigation.

For each CI test run, the tool records detailed screencasts at the native frame rate of the device, typically 60 frames per second, allowing for precise detection of even subtle lags. The recorded data includes timestamps for each frame, system resource usage (e.g., CPU and memory consumption), and any detected lag instances classified into three categories: janky frames, long loading frames, and frozen frames. By capturing this comprehensive dataset, we can analyze the occurrence, duration, and frequency of different types of GUI lags in a variety of testing scenarios.

\subsubsection{Labeling Ground Truth Dataset} 
To create the ground truth dataset to evaluate \tool, we involved five user experience experts from Company A. These experts were selected based on their extensive experience in mobile application testing. 
Each expert independently watched the screencast playback and reviewed the corresponding images to identify lags. They assessed the screencast to determine if the lag was noticeable enough to affect the user experience, marking the exact start and end frames for each instance where a GUI lag was observed. 
Any GUI lag instance identified by at least one expert was classified as an issue. This approach allowed us to capture even subtle lags that may not affect all users equally but could still detract from the overall user experience. For each detected lag, the experts marked the relevant frames and provided details on the perceived severity based on both the type of lag and its duration.

\subsection{Accuracy of \tool in Detecting GUI Lags}


We use the dataset collected and labeled by the user experience experts as the ground truth for evaluation. 
In the context of a screencast, an instance GUI lag is considered correctly located by \tool if and only if it matches a lag (annotated as $[f_{start}, f_{end}]$) in the ground truth, with identical start and end frame indices (i.e, the same $f_{start}$ and $f_{end}$ values). 
To evaluate the effectiveness of \tool in detecting GUI lags, We employ precision, recall, and F1-score:

\begin{itemize}
    \item \textbf{Precision} is the fraction of correctly identified GUI lags out of all instances that \tool identified as GUI lags. It is defined as: 
\end{itemize}
\begin{equation}
Precision = \frac{\#\textit{Correctly identified GUI lags}}{\#\textit{Identified GUI lags}}
\end{equation}

\begin{itemize}
    \item \textbf{Recall} is the fraction of correctly identified GUI lags out of the total number of actual GUI lags. It is defined as: 
\end{itemize}
\begin{equation}
Recall = \frac{\#\textit{Correctly Identified GUI lags}}{\#\textit{Actual GUI lags}}
\end{equation}

\begin{itemize}
    \item The \textbf{F1-score} is the harmonic mean of precision and recall:
\end{itemize}
\begin{equation}
F_{1} = 2 \times \frac{Precision \times Recall}{Precision + Recall}
\end{equation}

Table~\ref{tab:resutls_detecting_lags} presents the results of detecting GUI performance lags using \tool, demonstrating its high level of accuracy in detecting these lags. The precision for detecting the three types of GUI lags ranges from 0.91 to 0.92, while recall ranges from 0.95 to 0.98, and the F1-score ranges from 0.93 to 0.95. Overall, \tool achieves high average precision of 0.91 and recall of 0.96 across all lag types. 

\begin{table}
	\centering
	\caption{Results of \tool in detecting performance GUI lags across different lag types.}
	\label{tab:resutls_detecting_lags}
	\scalebox{1}{
	\begin{tabular}{lrrr}
		\toprule
		Lag type&\tabincell{r}{Precision}&\tabincell{r}{Recall}&\tabincell{r}{F1-Score}\\
		\midrule
			Janky frames          &0.92	        &0.98       &0.95\\
   	  Long Loading frames   &0.91 		  &0.96		  &0.93\\
			Frozen frames         &0.91		    &0.95	    &0.93\\
		\midrule
		\tabincell{l}{Average Across Types} 	&0.91		&0.96 		&0.94\\
		\bottomrule
	\end{tabular}}
        \vspace{-5mm}
\end{table}

Even though \tool achieves high precision and recall, there are still some missed or incorrectly detected cases. Hence, we conduct further analysis on such cases. We found that certain factors contribute to these inaccuracies. Primarily, our approach uses a threshold of 100 milliseconds based on prior HCI studies~\cite{1968_AFIPS_Response_time_in_man_computer, 1994_Usability_Engineering} to determine the shortest duration that users can typically perceive a lag caused by janky frames. While this threshold works well in most scenarios, it is still possible to lead to missed detection or false positives. 
For example, janky frames with a duration under 100 milliseconds may still be perceptible when they occur in scenarios with frequent or complex screen changes, such as during video playback or fast-paced animations. In these cases, even slight irregularities become noticeable to users, but \tool may not flag them due to the short duration. Similarly, certain long loading frames and frozen frames that exceed the thresholds might not be detected as high-severity issues when users have context-specific expectations of delay, such as when waiting for large media files or complex images to load. In these instances, users anticipate the delay, which reduces their perception of it as a performance issue, yet \tool might flag it as a severe lag based on the duration alone.
This analysis indicates that to further improve detection accuracy, \tool needs to go beyond a simple time threshold and incorporate additional factors such as the type of user interaction, the visual prominence of the lagging element, and the expected behavior of certain app functionalities. By integrating these contextual cues, \tool can more effectively capture the GUI lags that are most likely to affect user satisfaction.

\rqbox{We find that \tool achieves an average precision and recall of 0.91 and 0.96 on detecting GUI lags from screencasts. Our analysis finds that incorporating additional app context may further improve detection results. }

		
		

		
\subsection{Reporting Performance Bugs Based on \tool's Results}
We applied the \tool to screencasts recorded during real-world mobile app testing, successfully identifying many instances of severe GUI lags with high precision and recall. 
These detected lag instances are particularly valuable for reporting performance bugs, as they indicate potential issues and provide developers with actionable insights for debugging and locating performance problems. Below, we outline how we prioritize the severe lags identified by \tool and generate corresponding performance bug reports. 


We prioritize the identified severe lags based on their frequency, duration, user interaction, and visual context to ensure effective ranking, as higher-ranked lags are more likely to be perceived as performance issues. 1) \textbf{Frequency:} Each severe lag is assigned a rank level, with more frequent occurrences receiving a higher rank. For example, janky frames that occur repeatedly are flagged as a high-priority issue due to their potential to disrupt user interactions consistently. 2) \textbf{Duration:} The longer the duration of the severe lags, the higher their rank. For example, load frames that persist for extended periods are prioritized over those with shorter duration. 3) \textbf{User Interaction:} Lags during high-interaction events receive higher priority. Users are less likely to tolerate delays in response to direct interactions, such as scrolling. 
4) \textbf{Visual Context:} We also consider the visual prominence of each lagging element. Loading frames affecting a major visual component on the screen, such as the main content area, are prioritized over those impacting minor or secondary elements. 



After prioritizing the identified severe lags, we generate comprehensive bug reports that include the following information:
\begin{itemize}
    \item Test Scenario: The specific test case and user interaction that trigger the lag.
    \item Type and Ranking: The lag type (janky, long loading, or frozen frames) along with its ranking.
    \item Frequency and Duration: Metrics on how often the lag occurred and its duration, helping developers to understand the impact and consistency of the issue.
    \item Context: Information on where the lag occurred, including the start and end frames, the user interaction involved, and the affected element.
    \item System and Resource Data: Performance metrics such as CPU and memory usage during the lag to assist in diagnosing underlying causes.
    \item Screencast Segment: A video snippet of the affected frames, enabling developers to visually assess the lag's impact on user experience.
\end{itemize}




We received positive feedback from the developers on the performance bug reports based on the lags. They confirmed that the top-ranked bug reports have a larger impact on user satisfaction. Thus, ranked bug reports help developers focus on the most user-centric issues. The comprehensive information provided, such as the specific test scenario, user interactions, and the frequency of the lags, greatly assists developers in reproducing bugs and identifying their root cause.

\rqbox{Positive practitioner feedback confirms that the comprehensive bug reports generated from the lags detected by \tool help developers focus on the more critical issues and debug them efficiently.}

\section{Production Impact and Discussion}
\label{sec:discussion}
\begin{figure*}
	\centering
    \includegraphics[width=0.8\textwidth]{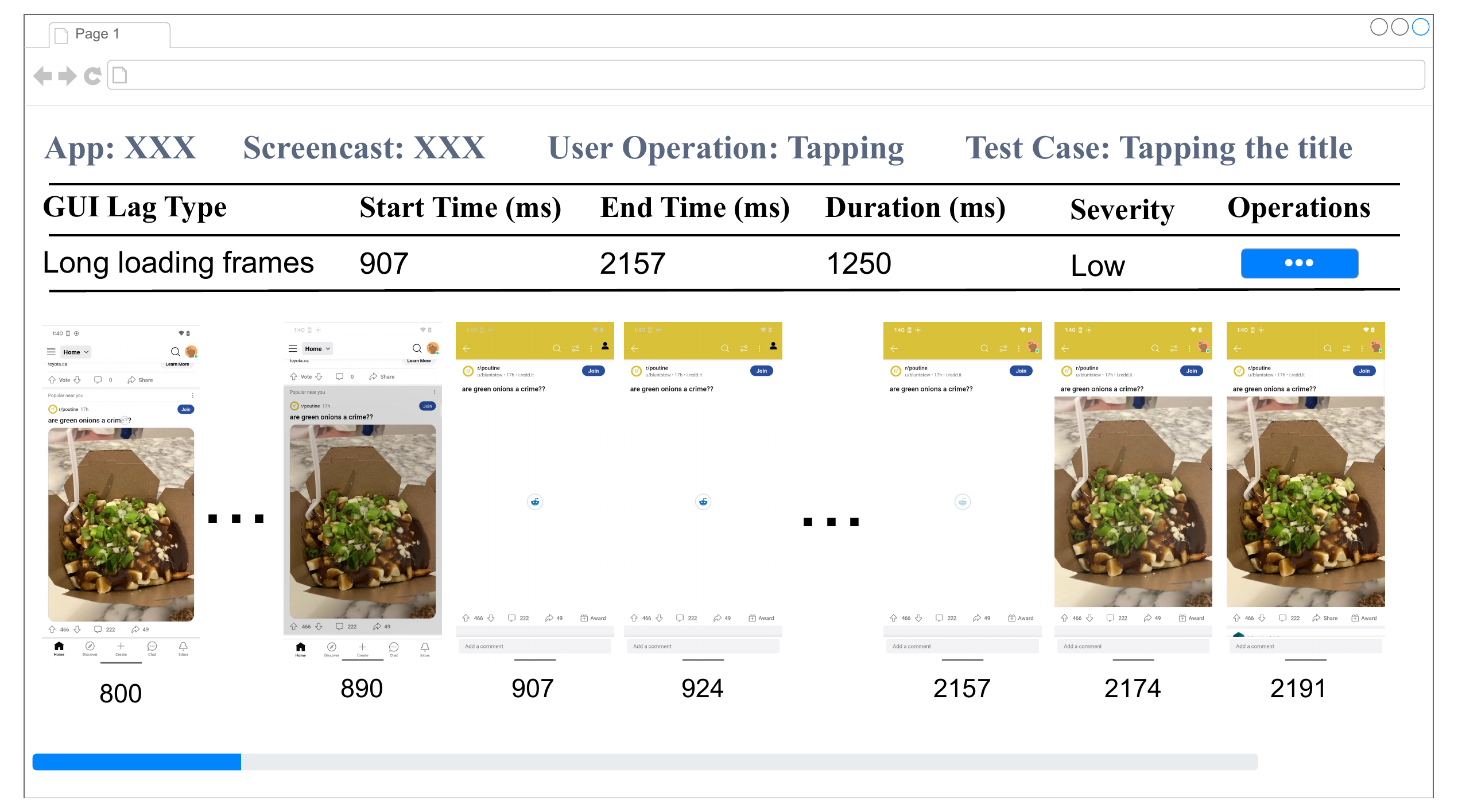}
	\caption{An illustrative interface of \tool. Due to the NDA, we cannot show the actual interface of \tool. However, to better showcase \tool, we created an illustration of how the result may be presented. \tool's report shows the detected GUI lags, their corresponding types, start and end times (and the frames), duration, and severity.}
	\label{fig_interface}
\end{figure*}
Figure~\ref{fig_interface} provides an illustration of \tool's result presentation. Due to the NDA, we cannot show the actual user interface of \tool, but we hope to provide an illustrative example. \tool's report shows the detected GUI lags, their corresponding types, start and end times (and the frames), duration, and severity. The detection report is linked to the screencast and also shows the user interaction of the tests when recording the screencast. Once a user wants to investigate one specific GUI lag, \tool shows the recorded screencast and the corresponding frames that experience a lag. This process significantly facilitates the analysis. We received much positive feedback from practitioners after deploying \tool in production. Below, we discuss the feedback we received that we hope can inspire other practitioners and future research. 

\subsection{Effectiveness of \tool in Detecting User-Centric Issues}
\tool has been consistently operating online for a while. With each code update, we execute automated GUI tests on the mobile device and record the resulting screencasts, which are then analyzed by \tool. In our evaluation, \tool successfully identified critical instances of GUI lag, most of which have been validated by developers in follow-up assessments. This capability is invaluable in identifying user-experienced performance issues that conventional metrics, such as CPU or memory usage, often fail to capture. 
Unlike traditional metrics-based tools, \tool's focus on perceptible lag types (i.e., janky, long loading, and frozen frames) ensures that it captures issues aligned with what users actually notice during interactions. Developers have noted that this capability helps them address performance issues that conventional tools might overlook, particularly those that influence user perception and satisfaction. For instance, \tool has proven effective in highlighting minor yet impactful lags that occur during high-interaction events, such as tapping or scrolling, which can lead to noticeable disruptions if left unaddressed. This user-centric approach to performance monitoring has proven instrumental in helping teams prioritize fixes based on real-world impact, making \tool an essential tool for enhancing the responsiveness and fluidity of mobile applications.

\subsection{Usefulness of \tool}
\phead{\tool Enhances Debugging Efficiency.} 
\tool significantly enhances debugging efficiency by providing comprehensive information about detected lags, helping developers identify potential performance issues. \tool accurately indicates the locations and types of GUI lags within screencasts, user operations (e.g., tapping, swiping, scrolling), and specific test scenarios, offering insights into how these lags occur. By focusing on user-perceived lags, developers can better understand how these lags directly impact users. Additionally, prioritizing these lags allows developers to concentrate on the most critical user experience issues. 
As one practitioner stated, ``\textit{The comprehensive reports have drastically reduced our debugging time and helped us focus on the critical performance issues that matter most to our users.}''

\phead{\tool Enhances Continuous Performance Monitoring.}
\tool is vital in monitoring mobile app performance from the GUI perspective. By testing different versions of an application and examining the distribution of detected GUI lags, developers can identify performance degradation and evaluate whether user-centric performance has declined over time. Additionally, \tool’s integration with Continuous Integration (CI) pipelines enables automated GUI performance monitoring with every build, helping teams catch regressions early in the development cycle. As one practitioner noted, ``\textit{\tool has become an integral part of our CI pipeline. Its reports allow us to quickly identify and address new performance lags introduced by recent changes, enabling us to deliver smoother experiences with each release.}''
\tool's regular reporting helps practitioners track performance trends, identify recurring issues, and proactively address potential bottlenecks before they affect users.

\subsection{Flexibility of \tool}
\tool demonstrates strong adaptability in both \textit{GUI Testing} and \textit{Lag Detection}. It supports mobile screencasts recorded from various testing methods, whether through manual testing or automated GUI testing tools, ensuring compatibility with diverse testing practices.
Furthermore, \tool is highly adaptable across multiple platforms and GUI applications, including Android, iOS, and TV. Minimal adjustments, such as fine-tuning detection thresholds, are required for it to work effectively across platforms. Currently, we are working closely with practitioners to adapt \tool to other UI-centric testing on various platforms.  


\section{Related Work}
\label{sec:related}

In this section, we discuss the works related to our paper.

\subsection{Empirical studies on Mobile Performance}

Several empirical studies~\cite{ 2014_ICSE_Characterizing_and_detecting_performance_bugs, 2023_empirical_study_on_mobile_performance, 2020_EMSE_statically_detectable_performance_issues} have investigated and categorized performance issues in mobile applications.
Liu et al.~\cite{2014_ICSE_Characterizing_and_detecting_performance_bugs} categorized performance issues into three main categories: GUI lag, memory bloat, and energy leaks. Notably, GUI lag accounted for over 75\% of the performance issues, significantly reducing app responsiveness.
Rua et al.~\cite{2023_empirical_study_on_mobile_performance} examined how semantic code changes impact performance metrics like energy usage, runtime, and memory consumption. Their findings revealed inconsistencies between these metrics and the priorities set by Android Lint. 
These studies highlight the need for more advanced tools and techniques to accurately detect and address performance issues in mobile applications, ultimately improving user experience and app reliability.

\subsection{Detecting Mobile Performance Issues}
Many studies have used static analysis of source code to detect performance issues in mobile applications~\cite{2014_ICSE_Characterizing_and_detecting_performance_bugs, 2019_EMSE_iPerfDetector,2019_SANER_Characterizing_and_Detecting_Inefficient_Image_Displaying_Issues, 2020_EMSE_statically_detectable_performance_issues, 2021_ICSE_IMGDroid_Detecting_Image_Loading_Defects, 2023_ASE_Detection_Thread_Misuses}. Specifically, some studies \cite{2014_ICSE_Characterizing_and_detecting_performance_bugs, 2019_EMSE_iPerfDetector, 2023_ASE_Detection_Thread_Misuses} focus on statically analyzing the source code to identify performance issues based on performance patterns, while others \cite{2019_SANER_Characterizing_and_Detecting_Inefficient_Image_Displaying_Issues, 2021_ICSE_IMGDroid_Detecting_Image_Loading_Defects} target inefficient image displaying issues. Static analysis tools like Android Lint~\cite{Android_Lint}, FindBugs~\cite{FindBugs}, PMD~\cite{PMD} are also used for detecting performance issues.
However, the issues identified through static analysis may not accurately reflect the user's experiences of app unresponsiveness. Moreover, static analysis tools rely heavily on known performance patterns, reducing their effectiveness in real-world scenarios.

Other profiling tools, such as Android Debug Bridge (adb)~\cite{adb} and Perfetto~\cite{perfetto}, monitor  system metrics like CPU and memory usage to detect performance issues. While these metrics provide valuable insights into system performance, they do not always reflect the responsiveness of the user interface (UI)~\cite{2022_IST_resource_influences_UI_responsiveness}, making it difficult to identify user-perceived issues such as GUI lags. 
Researchers have also developed dynamic instrumentation tools like AppInsight~\cite{2012_OSDI_AppInsight}, PerfProbe~\cite{2019_MOBILESoft_PerfProbe}, AppSPIN~\cite{2022_EMSE_AppSPIN} to collect system performance metrics and program events (e.g., user inputs, network events, UI thread activity) to detect performance issue. They still struggle to capture the full scope of user-perceived responsiveness. In contrast, our approach analyzes mobile screencasts, which provide direct insight into how applications behave from the end user’s perspective, making them ideal for identifying performance issue, particularly those related to UI responsiveness.

\subsection{Computer Vision in Mobile Testing}
As computer vision techniques evolve, they have been applied to enhance the efficiency and accuracy of mobile app testing. Some studies have employed these techniques to support testing by improving exploration efficiency~\cite{2023_ICSE_Efficiency_Matters_Speeding_Up_Automated_Testing_with_GUI_Rendering_Inference}, recording and replaying tests~\cite{2020_ICSE_translating_video_recordings_of_mobile_app_usages, 2023_UIST_Video2Action}, and enabling cross-platform testing~\cite{2024_TOSEM_PIRLTEST_GUI_Testing_via_Image_Embedding_and_RL, 2021_ICSE_Layout_and_Image_Recognition_Driving_Mobile_Testing}. In contrast, our approach focuses on analyzing the results (i.e., mobile screencasts) of GUI testing to detect performance issues, thus expanding the mobile testing.

Other studies have used computer vision techniques to analyze the mobile screenshot or screencast for detecting visual issues in mobile application GUIs. For example, some studies~\cite{2018_ICSE_Automated_reporting_of_GUI_design_violations, 2020_ICSE_Seenomaly_vision_based_linting_of_GUI_animation_effects_against_guidelines, 2021_ICSE_Hunting_Down_Visual_Design_Smells_in_Complex_UIs, 2024_ICSE_MotorEase} focus on identifying UI design issues by checking for compliance with design guidelines or intended designs. Additionally, other research addresses GUI-related bugs, such as UI display issues~\cite{2021_ASE_spotting_UI_display_issues}, data inconsistency bugs~\cite{2024_ICSE_Data_Inconsistency_Detection_of_Mobile_Apps}, and text glitches in game app GUIs~\cite{2023_FSE_Automated_Game_GUI_Text_Glitch_Detection}. In contrast, our work specifically applies computer vision techniques to analyze mobile screencasts to detect user-perceived GUI lags that affect user experiences.  

\section{Conclusion}
\label{sec:conclusion}


In this paper, we presented \tool, a novel framework designed to detect GUI lags in mobile applications. \tool identifies user-perceived performance issues, such as janky frames, long loading frames, and frozen frames, by analyzing screencasts directly from the end user perspective. Unlike traditional performance analysis tools that depend on system-level metrics, \tool offers a user-centered approach that aligns more closely with real-world user experiences. 
Through our collaboration with Company A, we successfully integrated \tool into a production environment, enabling continuous monitoring of GUI performance as part of their CI pipeline. The evaluation demonstrated \tool's high precision and recall. We also discuss how \tool helps developers prioritize fixes by providing detailed information on each lag's type, duration, and context.  
We hope this work can inspire practitioners and researchers to adopt more user-centered approaches in their performance testing processes. 


\section{Acknowledgement}
\label{sec:ack}
We want to thank Company A for providing access to the enterprise systems we used in our case study. The findings and opinions expressed in this paper are those of the authors and do not necessarily represent or reflect those of Company A and/or its subsidiaries and affiliation. Our results do not in any way reflect the quality of Company A's products.

\balance
\footnotesize
\bibliographystyle{IEEEtranN}
\bibliography{IEEEabrv,ref}

\end{document}